\documentclass[runningheads,hidelinks]{llncs}

\pdfoutput=1

\usepackage[space]{cite}
\usepackage{amsmath,amssymb,amsfonts}
\usepackage{algorithmic}
\usepackage{graphicx}
\usepackage{textcomp}
\usepackage{xcolor}
\usepackage[utf8]{inputenc} 
\usepackage{listings}
\usepackage{booktabs}
\usepackage{float}
\usepackage{url}
\usepackage{microtype}
\usepackage[subtle]{savetrees}
\usepackage[caption=false]{subfig}
\usepackage[hidelinks]{hyperref}

\def\BibTeX{{\rm B\kern-.05em{\sc i\kern-.025em b}\kern-.08em
    T\kern-.1667em\lower.7ex\hbox{E}\kern-.125emX}}

\definecolor{cloudwhite}{cmyk}{0,0,0,0.025}
\definecolor{deepblue}{rgb}{0,0,0.5}
\definecolor{deepred}{rgb}{0.6,0,0}
\definecolor{deepgreen}{rgb}{0.21,0.56,0.42}

\newcommand{\mysubsubsection}[1]{\vspace{0.15cm}\noindent\textbf{#1}\,}

\begin{document}

\title{Determining Microservice Boundaries:\\ A Case Study Using Static and Dynamic\\ Software Analysis}

\titlerunning{Determining Microservice Boundaries}
\author{Tiago Matias\inst{1} \and
Filipe F. Correia\inst{1,2} \and
Jonas Fritzsch\inst{4,5} \and
Justus Bogner\inst{5,4} \and\\
Hugo S. Ferreira\inst{1,2} \and
André Restivo\inst{1,3}}
\authorrunning{T. Matias et al.}
\institute{Faculty of Engineering, University of Porto, Portugal \and
INESC TEC, FEUP Campus, Portugal \and
LIACC, FEUP Campus, Portugal \and
Institute of Software Technology, University of Stuttgart, Germany \and
University of Applied Sciences Reutlingen, Germany\\[0.1cm]
\email{\{up201700421,filipe.correia,hugosf,arestivo\}@fe.up.pt}\\
\email{\{jonas.fritzsch,justus.bogner\}@iste.uni-stuttgart.de}
}
\maketitle              %
\begin{abstract}

A number of approaches have been proposed to identify service boundaries when decomposing a monolith to microservices. However, only a few use systematic methods and have been demonstrated with replicable empirical studies.
We describe a systematic approach for refactoring systems to microservice architectures that uses static analysis to determine the system's structure and dynamic analysis to understand its actual behavior. A prototype of a tool was built using this approach (MonoBreaker) and was used to conduct a case study on a real-world software project.
The goal was to assess the feasibility and benefits of a systematic approach to decomposition that combines static and dynamic analysis. The three study participants regarded as positive the decomposition proposed by our tool, and considered that it showed improvements over approaches that rely only on static analysis.

\end{abstract}

\section{Introduction}\vspace{-0.2cm}

The \textit{microservices} architecture steadily gained popularity over the last years. Nowadays, it is often used in greenfield projects, but a lot of the times, systems are first developed as monoliths, which are quicker to develop and to test than microservices. Monoliths can then be broken up into microservices, when and if the need arises~\cite{monolith-first}. Doing this may promise high scalability, shorter release cycles or better maintainability. However, missing to identify the right boundaries may hinder reaching these benefits~\cite{Balalaie2016}. Therefore, an essential part of such a refactoring is the decomposition approach~\cite{Fritzsch2019a}, which has the end-goal to identify contextually-related functionality and encapsulate it into different services. These should be characterized by a high cohesion inwards and loose coupling outwards. To optimally leverage from the microservices architectural pattern, existing functionality has to be split up with appropriate granularity as well.

There have been a number of approaches proposed already to decompose monoliths into microservices~\cite{Ponce2019,Fritzsch2019}. However, Fritzsch et al. found that such refactoring approaches were often not considered by practitioners and that identifying suitable service cuts is still perceived as a major challenge~\cite{Fritzsch2019a}. They asked 16 practitioners from 10 companies who were in the process of migrating their systems. Participants were either not aware of such tools or even convinced that it would be impossible to automate such a complex task. In a review of refactoring approaches, the same authors ascribed a lack of automation and missing tool support to most approaches proposed by academia~\cite{Fritzsch2019}. This lack of tools inhibits adoption in industrial contexts and makes empirical studies more challenging to conduct.

We address this gap by \textit{a)}~identifying a systematic approach that combines principles of the previously proposed methods, \textit{b)}~using it to create a prototypical implementation, and \textit{c)}~conducting an industry case study with the prototype.

In the remainder of the paper, we discuss \textbf{related work} to provide an overview of other approaches, and describe our own \textbf{approach}, which relies on static and dynamic analysis. We introduce a \textbf{prototype}  -- MonoBreaker -- that embodies this approach, and that identifies service boundaries for monoliths based on the Django web framework. Afterward, we present a \textbf{case study}, in which we contrast the results of MonoBreaker with ServiceCutter by surveying three developers of the project.

\vspace{-0.1cm}\section{Related Work}\vspace{-0.1cm}
\label{sec:relatedwork}

The subject of decomposing and migrating monolithic applications to microservices is addressed in books such as \emph{Building Microservices}~\cite{building-microservices}, \emph{Monolith to Microservices}~\cite{mono-to-micro} and \emph{Microservices Patterns}~\cite{RichardsonBook2018}. Likewise, a variety of research papers describe ways to tackle such transformations. 

Building microservices ideally means to create services that are highly cohesive and loosely coupled. Tyszberowicz~\cite{Tyszberowicz2018} confirms that Domain-Driven Design (DDD) is the most common technique for modeling microservices. With DDD, the software mirrors business domains and sub-domains as well as the related domain models and bounded contexts. Each bounded context implements a small set of strongly-related behaviors and conforms to the Common Closure Principle~\cite{bobclean}. These sets of behaviors shape individual units, resulting in cohesive designs of loosely-coupled services~\cite{domain-driven-design-book}. A system following DDD supports a higher degree of team independence as well as better scalability, testability and changeability~\cite{millett2015}.

\mysubsubsection{Meta-studies}
Ponce et al. provide an up-to-date overview in their review of 20 papers of migration and refactoring techniques~\cite{Ponce2019}. Their study focuses on the approaches, the applicability to certain system types, validations of the techniques, and the associated challenges. The authors group works by their underlying decomposition approaches: \textit{model-driven} (involving design elements, e.g., DDD), \textit{static analysis} (based on source code) and \textit{dynamic analysis} (based on runtime data).

Fritzsch et al. similarly compare 10 refactoring approaches and likewise provide a classification~\cite{Fritzsch2019}. They distinguish decompositions based on \textit{Static Code Analysis}, \textit{Meta-Data}, \textit{Workload-Data}, and \textit{Dynamic Microservice Composition}. While the first three classes imply a fixed decomposition result, a dynamic composition of services would be continuously re-calculated, e.g., based on workload constraints. The study moreover reveals that most approaches are only applicable to certain types of applications, require significant amounts of input, or have limited and prototypical tool support.

\mysubsubsection{Concrete Approaches and Tools}
Nunes et al. pursue an approach based on identifying transactional contexts of business applications and using a clustering algorithm to determine service candidates~\cite{Nunes2019}. Chen et al. similarly base the decomposition on the data flow of the business logic~\cite{Chen2018}. They compare the resulting service cut with the output of ServiceCutter~\cite{service-cutter-site}, a freely-available tool implementing the approach by Gysel et al.~\cite{Gysel2016}. ServiceCutter applies a clustering algorithm to identify new services and currently supports the \textit{Girvan-Newman} and \textit{Leung} algorithms for this purpose. To calculate the service cut, it requires that an \textit{Entity-Relationship Model} (ERM) of the system is given in a specific format along with \textit{User Representations} and \textit{Coupling Criteria}. The collection of these partly-exhaustive system specifications is done in a manual process and requires the help of domain experts.

Ren et al. acknowledge the inadequacy of approaches only relying on static analysis~\cite{Ren2018}. They recognize that not analyzing the runtime behavior would hinder the calculation of a complete and accurate service cut. Therefore, they combine static and dynamic analysis based on the applications' runtime behavior. A subsequent clustering calculates the candidate service cut. 
Likewise, Taibi et al. propose a combined approach based on dependency analysis and process mining techniques~\cite{Taibi2019}. The decomposition encompasses execution path and frequency analysis. After removing circular dependencies, additionally specified decomposition options are ranked based on coupling and granularity metrics to produce the candidate service cut. The authors employ a tool \footnote{More information is found at the tool's website -- \url{https://fluxicon.com/disco/}} that is capable of generating graphical visualizations to represent the business processes. Although a tool is referenced to capture the dynamic behavior of the system, the suggestion of service cuts is outside the scope of the work and must be done by experts, even if the authors mention that the process can somehow be automated.

\mysubsubsection{Implications for our Approach}
The methods described by Richardson \cite{RichardsonBook2018} (\textit{Decompose by business capability} and \textit{Decompose by subdomain}) provide general guidelines for a partly-automated decomposition process. They support architects in choosing appropriate input values and assessing the resulting candidate service cuts. 

The two meta-studies by Ponce and Fritzsch yield a variety of strategies to break down a monolith. Most do not combine static and dynamic analysis to steer the decomposition. As such, the works by Ren et al. and Taibi et al. comprise the core concepts of the approach described in our work. These works do not provide tools for service decomposition, or for any form of automation, but we will build on the concept of gathering runtime behavior and its analysis.

ServiceCutter is also of importance to our work, as it too implements the deterministic \textit{Girvan-Newman} algorithm. In some aspects, our work is less sophisticated than ServiceCutter, as it does not yet consider quality attributes like security, scalability, and business ownership. However, it trades that for the benefit of being independent from extraneous, subjective, information provided by experts to determine the service cuts.

\section{Approach}
\label{sec:approach}

Decomposing a monolith is often done based on insights from software developers on the specific context of the \textit{problem domain} and of the \textit{application's architecture}. The challenge that we aim to address is to reduce subjectivity, making the process more systematic and automated. The approach described below is based on ideas that have been documented before~\cite{Gysel2016,Ren2018,Taibi2019}, but are employed here for determining service boundaries with minimal to no manual input, which so far has not been feasibly demonstrated. Therefore, the approach is described as a hypothesis, and the case study in Section~\ref{sec:case_study} as a first step to provide support for its effectiveness.

In more concrete terms, this approach aims to be data-driven and to be independent of sophisticated input from experts. To do this, we do not take into account all the intricacies of the process as it is often done manually today. Instead, we focus on what information can be obtained from the application itself via \textit{static} and \textit{dynamic} analysis to find beneficial service cuts. We rely on the availability of \textit{a)}~static software artifacts, namely source code, and \textit{b)}~operational data, such as the use of API endpoints, of datastores, and of issued method calls.

\mysubsubsection{Static Analysis}
Software artifacts are analyzed and the collected information used to build a graph-like model of the system, representing components as \textit{nodes} and the dependencies between them as \textit{edges}. Components and dependencies can be of different types, and identifying them will depend on the used programming languages, frameworks and environments. For example, components can refer to \textit{classes}, \textit{packages} or \textit{modules}, and dependencies to \textit{imports} or \textit{method calls}. 

Each edge is assigned a \textit{weight} to represent the strength of the dependency. This is a function of the number and quality of connections between the two components. The weight of edges after static analysis can, for example, be the sum of the number of imports and method calls between its two components. 

\mysubsubsection{Dynamic Analysis}
The system is then monitored at runtime to gather \textit{operational data}, which is analyzed to identify how the dependencies are exercised during execution, and gain an understanding of how the system is actually used. Such information is used to compute a new weight for each edge of the graph. The \textit{final weight} values are a function of the \textit{static} and \textit{dynamic weights}, and are a measure for how the components in the system are mutually bound. The underlying assumption is that a high amount of interaction between two components correlates with belonging to a common bounded context. Including them in different microservices would imply higher costs in latency and in maintaining resilience and fault tolerance.

\mysubsubsection{Clustering}
A graph of the service composition will support identifying different clusters of components. The nodes connected by the edges with higher weight values will be grouped to form clusters of relatively \textit{high cohesion}. These clusters will depend on each other through edges with low weight values, representing relatively \textit{low coupling}. The clusters can, therefore, be used to determine a set of possible \textit{service cuts}. The specific clustering algorithm to be used is outside the scope of this approach, but would be interesting to explore (see Section \ref{sec:conclusions_and_future_work}).

\mysubsubsection{Decomposition Suggestion}
The identified service cuts serve as a foundation for assigning existing \textit{software artifacts} to each of the new services and advise on the architectural \textit{refactoring process}.\vspace*{-0.2cm}

\section{The MonoBreaker Tool}
\label{sec:prototype}

\begin{figure}[!hbtp]
  \vspace*{-0.3cm}  
  \centering
  \includegraphics[width=0.85\textwidth]{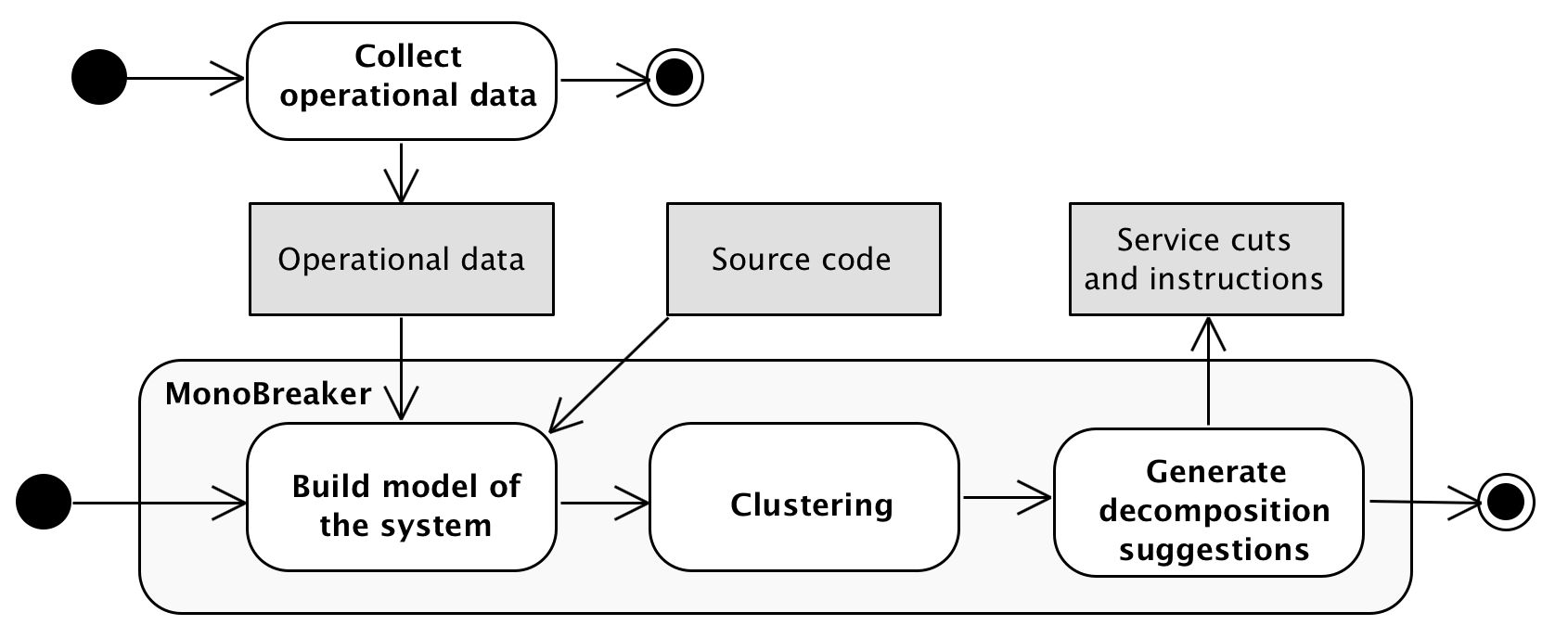}
  \vspace*{-0.3cm}
  \caption{Operation flow of MonoBreaker with inputs and outputs.}
  \label{fig:mono_flow}
  \vspace*{-0.8cm}
\end{figure}

\vspace{-0.1cm}MonoBreaker aims to demonstrate the feasibility of the approach and was used in the \textbf{case study} described in Section~\ref{sec:case_study}. It is a prototype\footnote{MonoBreaker is freely available at \url{https://github.com/tiagoCMatias/monoBreaker}.} and currently works with applications using the Django web framework. It takes a project's directory as input and does a \textbf{static analysis} of the source code to identify the overall project structure. This information is mapped to a graph-like model together with associated files and their dependencies. The same graph is populated with data collected through \textbf{dynamic analysis} to quantify the strength of the dependencies. The graph is then traversed to suggest a decomposition into new services, highlighting the source code files that will be involved and how the resulting services should communicate. This workflow is depicted in Figure~\ref{fig:mono_flow} and the several steps are exemplified below.

\subsection{Collect Operational Data}
\label{sec:collectoperationaldata}

\vspace{-0.1cm}Operational data is gathered using Silk, which is a profiling tool for Django\footnote{See \url{https://github.com/jazzband/django-silk} for more information.}. The tool is capable of supplying information about the usage of entrypoint methods (the ones invoked when a URL is requested), and the model classes and queries involved in the process of returning results from the database. It uses this information to infer some of the internal method calls, as we will see in the next section.

\subsection{Build Model of the System}
 
\vspace{-0.1cm}The static analysis inspects the domain model, the views, and the dependencies between them. In particular, it tracks the use of Django's \texttt{Model} class, identifying its subclasses (i.e., the \textit{domain model} of the application) and how they are connected through the declared \textit{foreign keys}. It also tracks the use of the \texttt{ModelViewSet} class by identifying its subclasses (i.e., the \textit{views} of the application) as well as the connections between these views and the model classes, via the \texttt{import} statements.

To illustrate the process, we present a minimalist example and the steps involved in suggesting a service decomposition using MonoBreaker. The file exemplified in Listing~\ref{lst:static_example} results in the extraction of the \texttt{ViewItem} class as a new graph node. The imports of \texttt{Attribute} (line~4) and \texttt{Item} (line~6) refer to subclasses of Django's \texttt{Model} class, therefore, these are also extracted as nodes, with graph edges connecting them to the \texttt{ViewItem} node. 
Both the model subclasses \texttt{Attribute} and \texttt{Item} have a connection to \texttt{ViewItem} because it imports them and invokes their methods. 

\begin{lstlisting}[
language=Python,
label={lst:static_example},
caption={The \texttt{ViewItem} class, an example of a \textit{view} in a Django application.}
]
from rest_framework.decorators import action
from rest_framework.response import Response
from rest_framework.viewsets import ModelViewSet
from ..models.Attribute import Attribute
from ..serializers.ItemSerializer import ItemSerializer
from ..models.Item import Item

class ViewItem(ModelViewSet):
  queryset = Item.objects.all()
  serializer_class = ItemSerializer

  @action(methods=['get'], detail=False)
  def get_item_details(self, request):
    if request.GET.get('attributes', None):
      data = self.serializer_class(self.queryset, many=True).to_representation(self.queryset)
      for item in data:
        item['attributes'] = Attribute.objects.get_by_item(item['id'])
      return Response(data)
    else:
      return Response(ItemSerializer(Item.objects.all(), many=True).data)
\end{lstlisting}

\begin{figure}[htb]
  \vspace{-0.2cm}
  \centering
  \subfloat[\label{fig:static_1}]{%
    \includegraphics[scale=0.63]{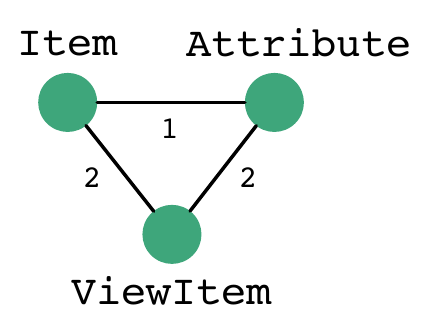}%
  }\hspace{0.1cm}
  \subfloat[\label{fig:mono_static_graph}]{%
    \includegraphics[scale=0.63]{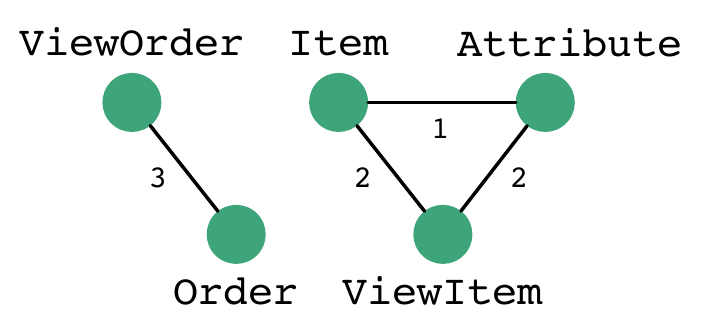}%
  }\hspace{0.1cm}
  \subfloat[\label{fig:final_graph}]{%
    \includegraphics[scale=0.63]{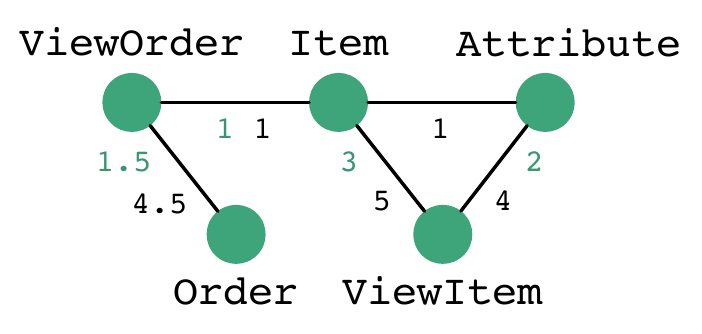}%
  }
  \caption{Each graph shows a different stage of the example, \protect\subref{fig:static_1}~is after analysing the \texttt{ViewItem} class, \protect\subref{fig:mono_static_graph}~after analysing the \texttt{ViewOrder} class, and \protect\subref{fig:mono_static_graph}~after incorporating the results from the dynamic analysis. Values in green are the weights determined by dynamic analysis alone, and those in black are the total weight produced up to that stage.}
  \label{fig:image2}
  \vspace{-0.3cm}
\end{figure}

Monobreaker uses the graph resulting from the analysis described thus far to generate the visual representation depicted by Figure~\ref{fig:static_1}. The \textit{weight} values associated to the edges represent the strength of the dependencies and are determined by:

\vspace{-0.4cm}\begin{align*}
StaticEdgeWeight = { NumImports } + { NumMethodCalls }
\end{align*}

\vspace{0.1cm}After the analysis of all source code files, a global dependency graph of the project is built. In this example, these files would also include Listing~\ref{lst:static_example_2}. 

\begin{lstlisting}[
language=Python,
label={lst:static_example_2},
caption={The \texttt{ViewOrder} class, an example of a \textit{view} in a Django application.}]
from rest_framework.decorators import action
from rest_framework.response import Response
from rest_framework.viewsets import ModelViewSet
from ..serializers.OrderSerializer import OrderSerializer
from ..models.Order import Order

class ViewOrder(ModelViewSet):
  queryset = Order.objects.all()
  serializer_class = OrderSerializer

  @action(methods=['get'], detail=False)
  def get_order_details(self, request):
    if request.GET.get('items', None):
      data = self.serializer_class(self.queryset, many=True).to_representation(self.queryset)
      for order in data:
        order['items'] = Order.objects.get_order_items(order['id'])
      return Response(data)
    else:
      return Response(OrderSerializer(self.queryset, many=True).data)

  def list(self):
    return Response(OrderSerializer(Order.objects.all(), many=True).data)
\end{lstlisting}

Figure~\ref{fig:mono_static_graph} represents the updated version of the graph after the static analysis of the second \textit{view} class. Note also the dependency between \texttt{ViewOrder} and \texttt{Item} via the call to the \texttt{get\_order\_items()} method. Detecting it could be attempted through deeper static analysis, in particular of chains of method calls that jump into framework code. The static detection of this dependency is a limitation of the current implementation of MonoBreaker, but it is one of little consequence, as it can still be detected through dynamic analysis, as we will see next.

The static analysis of the system is followed by the runtime analysis. The operational data that was previously collected (see Section~\ref{sec:collectoperationaldata}) is processed and the result used to update the graph with \textit{a)}~previously undetected dependencies (in this example, the one between \texttt{ViewOrder} and \texttt{Item}) and \textit{b)}~with updated weight values.
This ensures that we also consider the existence and the strength of dependencies that cannot be determined solely by inspecting the source code.

The requests received by the application may result in multiple method calls that eventually touch specific model classes. These are determined by MonoBreaker via the database queries that are issued during the processing of a specific request. Table~\ref{tab:operationaldata} shows some of the data resulting from the dynamic analysis, which is used to compute the dynamic weights.\vspace{-0.3cm}

\begin{table}[h!]
\centering
\begin{tabular}{ @{\hskip3pt}l @{\hskip6pt}l @{\hskip4pt}c @{\hskip6pt}l }
 \textbf{View} & \textbf{Method} & \textbf{\# Calls} & \textbf{Related Models}\\ 
\hline
 \texttt{ViewOrder} & \texttt{list()}                 & 2 & \texttt{Order} \\  
 \texttt{ViewOrder} & \texttt{get\_order\_details()}  & 4 & \texttt{Order}, \texttt{Item} \\  
 \texttt{ViewItem}  & \texttt{list()}                 & 4 & \texttt{Item} \\  
 \texttt{ViewItem}  & \texttt{get\_item\_details()}   & 8 & \texttt{Item}, \texttt{Attribute}\hspace{-0.1cm} \\
\hline\\
\end{tabular}
\caption{Data determined through dynamic analysis for this example.}
\label{tab:operationaldata}
\vspace*{-0.8cm}  
\end{table}

To keep the weight values calculated by the dynamic analysis in the same order of magnitude as those calculated from static analysis, MonoBreaker normalizes them -- the highest weight determined from the dynamic analysis will be at most as high as the highest one calculated from static analysis. Therefore, the equation representing the weight that arises from dynamic analysis becomes:\vspace{-0.5cm}

\begin{align*}
DynaEdgeWeight = NumMethodCalls \times \frac{MaxStaticWeight}{MaxNumMethodCalls}
\end{align*}

In this implementation, the weights from the static and dynamic analyses were considered in equal parts for determining the final weights, resulting in:\vspace{-0.4cm}

\begin{align*}
EdgeWeight = StaticEdgeWeight + DynaEdgeWeight
\end{align*}

Figure~\ref{fig:final_graph} depicts the resulting graph, showing the computed \textit{DynaEdgeWeight} in green and the final \textit{EdgeWeight} in black.

\subsection{Clustering}

The dependencies collected through the static and dynamic analyses are used by MonoBreaker to create a graph-like model of the system. Nodes consist mainly of Django \textit{model} and \textit{view} classes. A clustering algorithm is then applied to break the network down into smaller communities, thus grouping nodes according to the weights of the edges. We have chosen the Girvan-Newman algorithm\footnote{Connectivity-based clustering algorithm, such as Girvam-Newman, are based on the idea that nodes have more affinity to nearby nodes than to those farther way.}~\cite{girvan-newman-algorithm} given its apparent successful use in tools such as ServiceCutter. The resulting clusters indicate a set of potential service cuts.

\vspace{-0.2cm}\subsection{Generate Decomposition Suggestions}

After clustering the nodes, MonoBreaker provides an overview of the decomposition. It obtains the service cuts through the Girvan-Newman algorithm and provides the lists of the classes that will be needed for each service. These can be used by the developers to guide the refactoring process. Listing~\ref{lst:monobreaker_output} shows the output for our simple example.

\begin{lstlisting}[
language=Python,
label={lst:monobreaker_output},
caption={Example of an output of MonoBreaker.},
numbers=none
]
Total Files: 19
Django_Views: 2
Django_Models: 3

GraphNumber: 0
list_of_files: [
    'models.Attribute', 
    'models.Item', 
    'serializers.ItemSerializer', 
    'views.ViewItems'
]

GraphNumber: 1
list_of_files: [
    'models.Item', 
    'models.Order', 
    'serializers.OrderSerializer', 
    'views.ViewOrder'
]
\end{lstlisting}
\vspace{-0.3cm}

\subsection{Limitations}

The approach described in Section~\ref{sec:approach} is designed to apply to to a wide range of contexts. The tool described in this section, on the other hand, was designed with a narrower scope and it is worth highlighting some of its limitations.

\vspace{0.25cm}\noindent\textbf{Technologies} The opportunity, of using a fully developed monolith built with Django to conduct a case study in the industry, led us to develop MonoBreaker specifically for Django-based monoliths that use the object-relational mapper. At this point, the tool will work only for systems developed using these technologies.
    
\vspace{0.25cm}\noindent\textbf{Design assumptions} The implementation makes simplistic assumptions about the system to decompose, such as that it was designed around a domain model, and that it avoids cyclic dependencies and other kinds of unnecessary complexity. Such design problems should be approached before running MonoBreaker.
    
\vspace{0.25cm}\noindent\textbf{Operational time frame} The quality of the decomposition is sensible to the choice of an appropriate time frame for collecting operational data, as it should be representative of how the system is normally used. Functionality not used during the dynamic analysis time frame will not be considered for calculating dynamic weights.

\vspace{0.25cm}\noindent\textbf{Balancing quality attributes} Another assumption is that there is a single \textit{optimal} set of service cuts, but we know that there are often trade-offs when refactoring. Users of MonoBreaker are still not able to specify, for e.g., how the maintainability of the resulting system should be weighed against its scalability.

\section{Case Study} \label{sec:case_study}

To assess the feasibility and benefits of a systematic approach to decomposition that combines static and dynamic analysis, we conducted an industry case study using the developed prototype. We were interested in generating insights about the approach, in particular, in understanding its effectiveness for identifying good service boundaries when refactoring a monolith, and the impact that \textit{dynamic} analysis has on the decomposition result. For the latter part of the study, we turned to ServiceCutter for a comparison.

\subsection{Context}

The case study focused on a web application for supporting the collaboration between two centers of a logistics startup company. The application had 15 KLOC and more than 40 domain-model elements, and had recently gone through significant growth in its use, making it an interesting candidate for the study.

We achieved the participation of three of the four developers that form the team responsible for this application. Their professional experience was in the range of 1--5 years for two of the developers and 5--15 years for the third developer.

\subsection{Process}
\label{sec:process}

MonoBreaker was used to analyze the project and produced a suggestion for decomposing it into different services. The process consisted of four steps:
\begin{itemize}
    \item[a)] \textbf{Run MonoBreaker} -- We gathered the project source code and the runtime data collected through Silk and provided them as input to MonoBreaker, which used both \textit{static} and \textit{dynamic} analysis to produce a suggestion of how the system could be decomposed.
    
    \item[b)] \textbf{Run ServiceCutter} -- The data statically-collected in step \textbf{a)} was transformed to the ERM format expected by ServiceCutter and was provided as input to produce an alternative decomposition using \textit{static} analysis only.
    
    \item[c)] \textbf{Present MonoBreaker} -- A session was scheduled with the development team and included an introduction that explained the goal of the experiment and a showcase of MonoBreaker using an example project.
    
    \item[d)] \textbf{Questionnaire} -- Following the MonoBreaker demo, a questionnaire was handed out to the participants. It aimed to assess how the feasibility of the approach and the impact of \textit{dynamic} analysis on the quality of the results were perceived by the team. The participants did not have access to the source code during the questionnaire, and the two service decompositions were presented visually as dependency graphs. Participants were given 30 minutes to analyze the graphs and answer the questionnaire.
\end{itemize}{}

\vspace{-0.4cm}\subsection{Data Sources}

The case study used as data sources: \textit{a)}~the source code of the project, \textit{b)}~operational data collected through Silk during one week in a production environment and \textit{c)}~the answers to the questionnaire that were given by the team of the project.

The source code was obtained from the company's code repository. The operational information was collected in two tables created by Silk in the application's database (\texttt{silk\_request} and \texttt{silk\_sqlquery}). The questionnaire was built using Google Forms and the answers were gathered in a spreadsheet.

\vspace{-0.2cm}\subsection{Data Analysis}

Most questions were based on a Likert scale~\cite{Likert}, ranging from (1) \textit{Strongly Disagree} to (5) \textit{Strongly Agree}. Questions were organized into four groups. Below, we summarize the answers provided by the three interviewees for each group of questions. 

\mysubsubsection{Personal Experience}
These questions support understanding the team's professional experience, its familiarity with the case study project and with the process of migrating monoliths to microservices. The answers reveal that all team members have some experience migrating monoliths to microservices {\small(3, 4, 3)\footnote{Throughout this section, we'll use this notation to represent the answers of the three team members to a questionnaire item using a five-level Likert scale.}} and that they were very familiar with the case study project {\small(5, 5, 5)}, as expected. This ensures their ability to evaluate the decomposition approach.

\mysubsubsection{Approach}\label{sec:approachquestions}
The questions in this group aim to assess the perceived importance of different aspects when decomposing a monolith into microservices. If the understanding of these aspects by the study participants revealed to be different from our own, it could explain differences in the answers to questions in the next groups of questions. The questions and answers from the three developers are shown in Table~\ref{tab:questions-approach}. The results show unanimous agreement in that identifying the \textit{domain objects}, the \textit{relationships between components}, how these \textit{relationships are used in production} and the \textit{schema of the data store} are very important factors when determining potential new services {\small(5,5,5)}.

\begin{table}[H]
\renewcommand{\arraystretch}{1.2}
\begin{tabular}{p{0.875\linewidth}c}
\textbf{Question} \textit{[It's important ...]} & \textbf{Answers}\\
\toprule
\textit{... to know what methods are called between the components of the monolith} 						 	 & 5, 4, 3 \\
\midrule
\textit{... to know how frequently each method is called when the monolith is run in production}					 & 4, 3, 3 \\
\midrule
\textit{... to identify what the domain objects of the monolith are} 												 & 5, 5, 5 \\
\midrule
\textit{... to identifying what are the relationships between the monolith components}						 & 5, 5, 5 \\
\midrule
\textit{... to know how the relationships between the components are used when the monolith is run in production}	 & 5, 5, 5 \\
\midrule
\textit{... to identify what imports are made by each software component of the monolith} 							 & 5, 4, 3 \\
\midrule
\textit{... to identify what the schema of the database/datastore is} 												 & 5, 5, 5 \\
\midrule
\textit{... to know the operations made to the database/datastore} 								 & 5, 4, 4 \\
\midrule
\textit{... to identify how frequently the operations made to the database/datastore are executed when the monolith is run in production}      & 4, 3, 4\\
\bottomrule\\
\end{tabular}
\caption{Questions and answers in the \textit{approach} group.}
\label{tab:questions-approach}
\end{table}

\vspace{-0.6cm} The answers to the remaining questions were not unanimous, but still show that significant importance is attributed to knowing \textit{what operations are made to the database/datastore} {\small(5,4,4)}.

These results show the relevance, as perceived by the members of this team, of both structural and behavioral information for service decomposition, and therefore are aligned with the concepts that we used to define our approach.

\begin{figure*}[ht]
  \vspace{-0.2cm}
  \centering
  \includegraphics[width=0.85\textwidth]{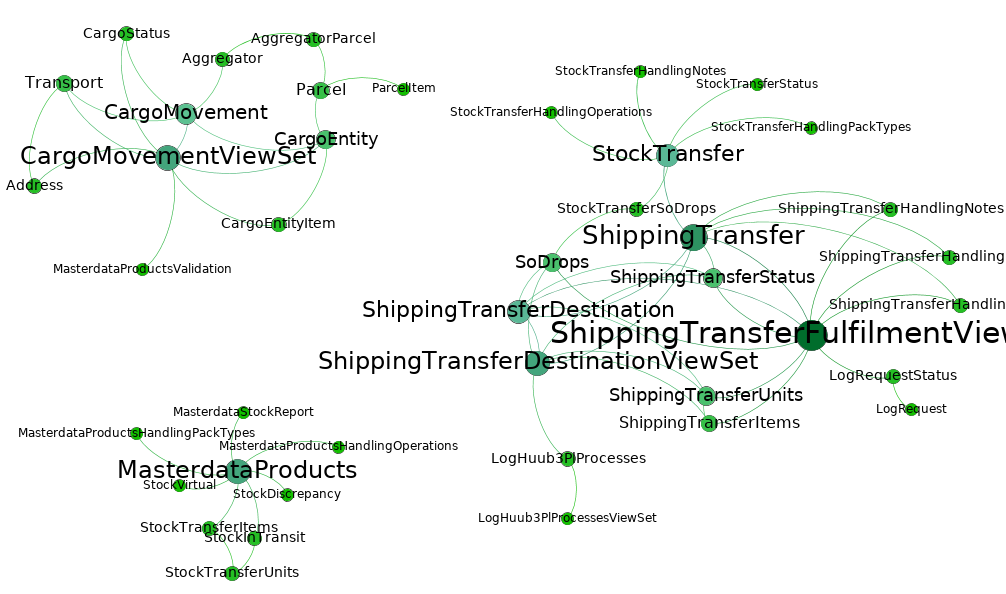}
  \vspace{-0.1cm}
  \caption{MonoBreaker decomposition result as depicted in the questionnaire.}
  \label{fig:breaker_output}
  \vspace*{-0.3cm}  
\end{figure*}

\mysubsubsection{Feasibility}\label{sec:feasibility}
The questions in this group evaluate the perceived feasibility of the approach regarding the quality attributes of the application. Namely, the questions focus on the scalability, ease of deployment, and ease of maintenance. They are supported by the decomposition created by MonoBreaker, which was visually presented as depicted by Figure~\ref{fig:breaker_output}. Both the questions and the answers are shown in Table~\ref{tab:questions-feasibility}.\vspace{-0.1cm}

\begin{table}[H]
\begin{tabular}{p{0.875\linewidth}c}
\textbf{Question} \textit{[The proposed decomposition as microservices ...]} & \textbf{Answers}\\
\toprule
\textit{... is the best one possible}													& 4, 3, 2 \\
\midrule
\textit{... is easier to scale (performance)}						& 4, 3, 2 \\
\midrule
\textit{... is easier to deploy new versions of the system}	& 4, 3, 2 \\
\midrule
\textit{... is easier for maintainability by the existing team(s)}	& 4, 3, 2 \\
\bottomrule\\
\end{tabular}
\caption{Questions and answers in the \textit{feasibility} group.}
\label{tab:questions-feasibility}
\vspace*{-0.4cm}
\end{table}

\vspace{-0.3cm}The participants did not agree in their answer to these questions but answered consistently to all the questions {\small(4,3,2)}. This led us to inspect more closely the answers for the \textit{justification} question (the open-ended question where they could provide further context to their answers) and conclude that the decomposition was perceived as a good basis, but insufficient. Namely, the decomposition consists of 3 services, but team members argued in favor of a more aggressive decomposition. Looking closely at Figure~\ref{fig:breaker_output}, we can see clusters around three different classes -- \texttt{CargoMovement}, \texttt{MasterdataProducts} and \texttt{ShippingTransfer}. From their answers, we understood that the team was expecting the \texttt{ShippingTransfer} cluster to be further decomposed into two distinct services. Section~\ref{sec:conclusions_and_future_work} outlines a few factors that can be explored in future work to improve the decomposition.

\mysubsubsection{Comparison With Using Only Static Analysis}
\label{sec:compstota}
This group has two Likert-scale questions, each accompanied by an open-ended \textit{justification} question. 

\begin{figure*}[ht]
  \centering
  \includegraphics[width=1.01\textwidth]{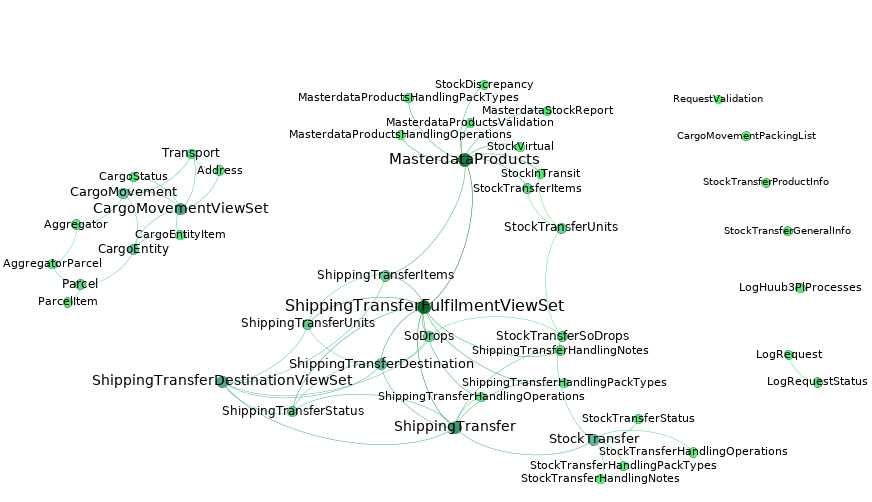}
  \caption{ServiceCutter decomposition result as depicted in the questionnaire.}
  \label{fig:serviceCutter_output}
  \vspace*{-0.4cm}
\end{figure*}

The first question compared the decomposition using both \textit{dynamic} and \textit{static} analysis with the one using only \textit{static} analysis. To ease the comparison between the outputs, we transported the information to Gephi\footnote{Gephi is a tool for graph analysis and visualization -- \url{https://gephi.org}.} and extracted both graphs. The graphs were depicted in the beginning of this group of questions as \textit{Decomposition A} and \textit{Decomposition B} (respectively, Figure~\ref{fig:serviceCutter_output} and Figure~\ref{fig:breaker_output}). 

The second question directly addressed the usefulness of the output provided by MonoBreaker, listing the classes that would be required by each service.

Table~\ref{tab:questions-comparingsota} shows the two questions and the associated answers.\vspace{-0.6cm}

\begin{table}[H]
\begin{tabular}{p{0.875\linewidth}c}
\textbf{Question} & \textbf{Answers}\\
\toprule
\textit{The decomposition A is better than the decomposition B} & 2, 2, 1 \\
\midrule
\textit{A tool to support decomposing a monolith into microservices would be useful if it provided this output} & 5, 5, 5 \\
\bottomrule\\
\end{tabular}
\vspace{-0.3cm}
\caption{Questions and answers in the "comparing with the state-of-the-art" group.}
\label{tab:questions-comparingsota}
\end{table}

The answers dismiss \textit{Decomposition A} as the best, concluding that combining \textit{static} and \textit{dynamic} analysis provided a better decomposition when compared to using \textit{static} only.

Regarding the output provided by MonoBreaker for guidance on the refactoring, the answers were unanimous in that it would be helpful.

\subsection{Threats to Validity}

The purpose of this case study is to gather evidence to support the approach. The design described in Section~\ref{sec:case_study} tries to minimize possible threats to validity, but those that exist need a closer look.

\mysubsubsection{Projects and participants} The sample of our case study was limited to one project and three software developers. The answers to the questionnaire's \textit{approach} group can be used to confirm if this team valued both structural and behavioral information when decomposing services, as these were base assumptions used to design our approach, but the small scale doesn't allow to generalize conclusions. We would certainly like to see this case study replicated for other products and larger organizations with different backgrounds, to verify if these preliminary results hold in other contexts.

\mysubsubsection{Possible biases from respondents} The partnership with the startup company for this case study was only possible due to good working and personal relationships and commitment between the company and the researchers. Therefore, there is always the possibility that the participants may have been inadvertently influenced. During the MonoBreaker presentation (Section~\ref{sec:process}), we took particular caution to take an impartial stance regarding the merits of the tool and of its underlying approach and to not interfere in any way when participants were responding to the questionnaire. Moreover, they didn't know which decomposition had been made using only \textit{static} analysis or using both \textit{static} and \textit{dynamic} analysis. For these reasons, we are confident in discarding this as a threat to validity.

\mysubsubsection{Representativeness of sampled data} The company supplied the project source code and allowed to alter it to enable the collection of operational data that otherwise would not be possible. As already mentioned, the operational data covered only one week of the application's run time information and collecting data for a longer period may have led to different results. All the relevant functionality of the application seems to have been used during this time, and we believe the amount of data to be sufficient to base a decomposition decision on. For this reason, we are confident in discarding this as a threat to validity.

\mysubsubsection{Suboptimal baseline} To assess the impact of \textit{dynamic} analysis in the decomposition, we compared the result of MonoBreaker (using \textit{static} and \textit{dynamic} analysis) with that of ServiceCutter (using \textit{static} analysis only). The choice of ServiceCutter stemmed from the intention to compare MonoBreaker with leading tools from the current state of the art. ServiceCutter is the only freely-available tool that we could run to automate the decomposition process with minimal manual input\footnote{This was possible by synthesizing a part of the inputs that it requires -- namely, the ERM -- and omitting the remaining inputs, which we were unable to create without resourcing to software developers -- namely, the \textit{User Representations} and the \textit{Coupling Criteria}.}.

However, we realized that the specific purpose of assessing the impact of dynamic analysis would have been better served by comparing the output of MonoBreaker when run with \textit{static} and \textit{dynamic} analysis with its output when run with \textit{static} analysis only. We believe that when the Girvan-Newman algorithm is chosen when running ServiceCutter, the resulting output should be identical to MonoBreaker's if only \textit{static} analysis is used, as MonoBreaker uses the same algorithm for clustering dependent components. Notwithstanding, running MonoBreaker with and without dynamic analysis would provide more robust evidence that no other factors had a significant influence on the decomposition result.

\section{Conclusions and Future Work}
\label{sec:conclusions_and_future_work}

In this work we contribute, \textit{a)} a systematic \textbf{approach} to decompose monolithic applications to microservices, \textit{b)} a \textbf{tool} prototype (MonoBreaker) that implements this approach and \textit{c)} the design and results of an industry \textbf{case study}.

The approach is based on previous ideas but differs in its focus on fully automating the process of determining service boundaries. It does so by relying on static and dynamic software analysis. The case study uses MonoBreaker to assess the feasibility and merits of the approach. The decomposition obtained by the tool was regarded positively by the participants and seen as an improvement over using only static analysis. MonoBreaker is freely available, and the methodological design is documented to enable the replication of the case study by other researchers.

\vspace{0.1cm}

To improve these contributions, several aspects will be addressed in future work:

\mysubsubsection{Model building} The approach doesn't define a specific way to build the model of the application using the results of \textit{static} and \textit{dynamic} analysis. Future work will evaluate if other algorithms for calculating the weight of dependencies may perform better than our current implementation, which is currently based on a set of simple heuristics.

\mysubsubsection{Clustering algorithms} The approach is also not prescriptive of a particular clustering algorithm. It will be interesting to evaluate if others render better results than Girvan-Newman, the one currently used by MonoBreaker.

\mysubsubsection{Evaluation metrics} To enable a more objective evaluation of the proposed decomposition, the approach could be extended with service-based metrics -- e.g., coupling and cohesion~\cite{Bogner2017}. The approach of Taibi et al.~\cite{Taibi2019} already includes metrics to rank decomposition candidates. A set of suitable service-based metrics for our approach would have to be determined, and can help to drive the search for better \textit{model-building} and \textit{clustering} algorithms.
    
\mysubsubsection{Comparison with human experts} Future studies will evaluate if a data-driven approach such as ours is, not only able to automate the decomposition process fully, but will also provide a better decomposition than human experts.
    
\mysubsubsection{Further studies} More industry case studies will need to be conducted to improve our understanding of the effectiveness and limitations of the approach, ideally with a diverse and significant number of applications and participants.
    
\mysubsubsection{Representativeness of sampled data} Future studies will compare the number of requests -- per request type -- that are received during the collection of operational data with those of more extended periods where operational data wasn't captured, but for which we are able to collect request statistics nonetheless. This will reinforce our confidence that the operational data collected is representative enough of a \textit{normal} use of the application.

\mysubsubsection{Fully automatic decomposition} MonoBreaker can identify file contents affected by the suggested decomposition, e.g., which class has to be extracted for each resulting service. The next step could be to suggest a sequence of lower-level refactorings required for the decomposition or even to automatically apply such refactorings to decompose the system.

\vspace{-0.1cm}\section*{Acknowledgment}
\vspace{-0.1cm}João Paiva Pinto and Isabel Azevedo discussed different forms of this work with us. We thank them for all the precious feedback.

\vspace{-0.2cm}\bibliographystyle{splncs04}
\bibliography{myrefs}

\end{document}